\documentclass[aps,prb,twocolumn,superscriptaddress,showpacs]{revtex4}

\usepackage{graphicx, bm, amsmath}

\newcommand{\dif}{\mathrm{d}}

\begin{document}


\title{NbSi nanowire quantum-phase-slip circuits: dc supercurrent blockade, microwave measurements and thermal analysis}

\author{C.\,H.\ Webster} \email{carol.webster@npl.co.uk}
\affiliation{National Physical Laboratory, Queens Road, Teddington, Middlesex, TW11 0LW, United Kingdom}

\author{J.\,C.\ Fenton} \email{j.fenton@ucl.ac.uk}
\affiliation{National Physical Laboratory, Queens Road, Teddington, Middlesex, TW11 0LW, United Kingdom}
\affiliation{London Centre for Nanotechnology, University College London, 17–-19 Gordon Street, London, WC1H 0AH, United Kingdom}

\author{T.\,T.\ Hongisto}
\affiliation{Physikalisch-Technische Bundesanstalt, Bundesallee 100, 38116 Braunschweig, Germany}

\author{S.\,P.\ Giblin}
\affiliation{National Physical Laboratory, Queens Road, Teddington, Middlesex, TW11 0LW, United Kingdom}

\author{A.\,B.\ Zorin}
\affiliation{Physikalisch-Technische Bundesanstalt, Bundesallee 100, 38116 Braunschweig, Germany}

\author{P.\,A.\ Warburton}
\affiliation{London Centre for Nanotechnology, University College London, 17–-19 Gordon Street, London, WC1H 0AH, United Kingdom}





\date{\today}

\begin{abstract}
We present a detailed report of microwave irradiation of
ultra-narrow superconducting nanowires. In our nanofabricated
circuits containing a superconducting NbSi nanowire, a dc blockade
of current flow was observed at low temperatures below a critical
voltage $V_{\rm c}$, a strong indicator of the existence of quantum
phase-slip (QPS) in the nanowire.  We describe the results of
applying microwaves to these samples, using a range of frequencies
and both continuous-wave and pulsed drive, in order to search for
dual Shapiro steps which would constitute an unambiguous
demonstration of quantum phase-slip. We observed no steps, and our
subsequent thermal analysis suggests that the electron temperature
in the series CrO resistors was significantly elevated above the
substrate temperature, resulting in sufficient Johnson noise to wash
out the steps.   To understand the system and inform future work, we
have constructed a numerical model of the dynamics of the circuit
for dc and ac bias (both continuous wave and pulsed drive signals)
in the presence of Johnson noise. Using this model, we outline
important design considerations for device and measurement
parameters which should be used in any future experiment to enable
the observation of dual Shapiro steps at experimentally accessible
temperatures and thus lead to the development of a QPS-based quantum
current standard.
\end{abstract}



\maketitle


\section{Introduction}
In recent years, there has been much interest in the phenomenon of
quantum phase-slip (QPS) in ultra-narrow superconducting nanowires.
This has been motivated by the potential applications for novel
devices based on QPS including a charge-insensitive
qubit\cite{Mooij-2005, Astafiev-2012} and a new quantum standard for
electrical current.\cite{Mooij-2006}

Research into other types of quantum current standard has also
intensified following proposals to redefine several of the SI base
units including the ampere.\cite{Milton-2007}  The tunable barrier
electron pump\cite{Giblin-2012} is the front-runner amongst the
competing technologies for realization of the ampere following
redefinition in terms of a fixed value of the charge on the
electron.\cite{Milton-2010} However, a quantum current standard
based on quantum phase-slip in a superconducting nanowire is an
attractive alternative, as it constitutes the exact dual of the
Josephson voltage standard and therefore offers the potential for
greater accuracy and scalability than the tunable barrier pump.  In
addition it would be a more elegant solution, being based on a truly
quantum effect, and offers the future possibility of combining
standards for current and voltage on a single
chip.\cite{Likharev-1987,Hriscu-2012}

It is highly desirable to carry out a direct metrological triangle
experiment in order to test the consistency of the present
understanding that $K_{\rm J} = 2e/h$, $R_{\rm K} = h/e^2$ and
$q_{\rm c}=2e$, where $K_{\rm J}$ is a constant characterizing the
Josephson voltage standard, $R_{\rm K}$ is a constant obtained from
the quantum Hall effect, and $q_{\rm c}$ is the charge transferred
per cycle of the drive signal in our quantum current standard.
 This requires that the output from a quantum
current standard should be accurate to better than 1 part in $10^7$
and that the magnitude of the current should exceed 100 pA to avoid
an excessively long integration time.\cite{Piquemal-2000,
Keller-2008, Milton-2010}  The present state of the art in GaAs
tunable barrier electron pumps is a current of magnitude 150 pA with
an experimentally demonstrated current accuracy better than 1.2
parts per million and evidence based on modelling that the true
accuracy approaches 1 part in $10^8$.\cite{Giblin-2012}

A quantum current standard based on QPS arises from the exact
duality between the Josephson junction and the quantum phase-slip
junction (QPSJ).\cite{Mooij-2006, Kerman-2012}  This leads to the
prediction that a QPSJ biased with a microwave signal should
generate quantised current steps in its current-voltage ($IV$)
characteristic, dual to the quantised-voltage Shapiro steps observed
in Josephson junctions which form the basis of the Josephson voltage
standard.  The QPSJ consists of an ultra-narrow superconducting
nanowire in series with an inductance and a resistance sufficiently
high for the single Cooper-pair regime to be reached
\cite{Mooij-2006, Kerman-2012} and overdamping of the circuit such
that charge fluctuations are suppressed.  The nanowire must have a
sufficiently high normal-state resistivity and sufficiently small
cross-sectional area to generate a sufficiently large QPS energy
$E_{\rm S}\gg k_{\rm B}T$ (dual to the Josephson energy $E_{\rm
J}=(\Phi_0/2\pi{})I_{\rm c}$ where $I_{\rm c}$ is the critical
current) and hence a sufficiently large critical voltage $V_{\rm
c}=2\pi{}E_{\rm S}/2e$ to enable a clear observation of dual Shapiro
steps.\cite{Arutyunov-review} Together with the constraint that the
superconducting critical temperature $T_{\rm c}$ should be
significantly higher than the lowest practical temperature
achievable in a laboratory, this places restrictions on the choice
of material for the nanowire.

At present, NbSi and InO$_x$ are considered to be the best
candidates.\cite{Fenton-2011, Astafiev-2012}  Ultra-narrow nanowires
have been developed in both these materials with effective diameters
in the range 20--40 nm.\cite{VanDerSar-2007, Hongisto-2012,
Johansson-2005}  QPS in InO$_x$ has been demonstrated via
spectroscopic measurements of $E_{\rm S}$ in a QPS
qubit,\cite{Astafiev-2012} while alternative evidence for QPS (in
fact, for the dual effect of coherent tunneling of single Cooper
pairs) has been observed in NbSi nanowires via quantum interference
in a device dual to the SQUID consisting of two QPSJs joined by a
superconducting island.\cite{Hongisto-2012} Preliminary signs of QPS
\cite{Lehtinen-2012b, Lehtinen-2012c} and of dual Shapiro steps
\cite{Lehtinen-2012} have been observed in a Ti nanowire, but only
traces of steps are observed.

In this article, we report dc and ac measurements of the $IV$ curve
of samples containing a NbSi nanowire embedded in a high impedance
environment ({\it i.e.},~a single QPSJ). We present measurements on
two samples, fabricated in the same facility as those reported in
Ref.~\onlinecite{Hongisto-2012}, and present representative results
here. The dc measurements show a blockade of Cooper-pair transport
along the nanowire, as expected with QPS; this is the dual of the
blockade of phase motion (with, therefore, no developed voltage) in
a Josephson junction for currents less than its critical current.
The ac measurements were performed to search for evidence of dual
Shapiro steps, predicted to occur at currents $I = 2enf$, where $f$
is the ac frequency and $n$ is an integer corresponding to the
number of Cooper pairs transferred per cycle. Observation of these
steps would constitute unambiguous demonstration of the QPS origin
of the effect and rule out other explanations of the dc results,
such as single-electron tunneling effects. However, no dual Shapiro
steps were observed; instead, as the microwave power was increased,
the blockade feature became less pronounced and increasingly
rounded. These observations are consistent with heating of the
sample. We describe thermal analysis and temperature-dependent
measurements of the $IV$ curve without microwave drive which
indicate that the electron temperature in the circuit was elevated,
resulting in Johnson noise sufficient to wash out any steps.

We also present a numerical model of the $IV$ curve for dc and ac
bias (both continuous wave and pulsed drive signals) in the presence
of Johnson noise.  We use this to predict an improved set of
experimental parameters, comprising resistors of larger volume and
higher resistivity and a pulsed drive signal with a higher pulse
amplitude and lower pulse width, to enable the observation at
experimentally accessible temperatures of dual Shapiro steps.

\section{Fabrication and circuit design}
\begin{figure}[!ht]
\begin{center}
\includegraphics[width=8.6cm]{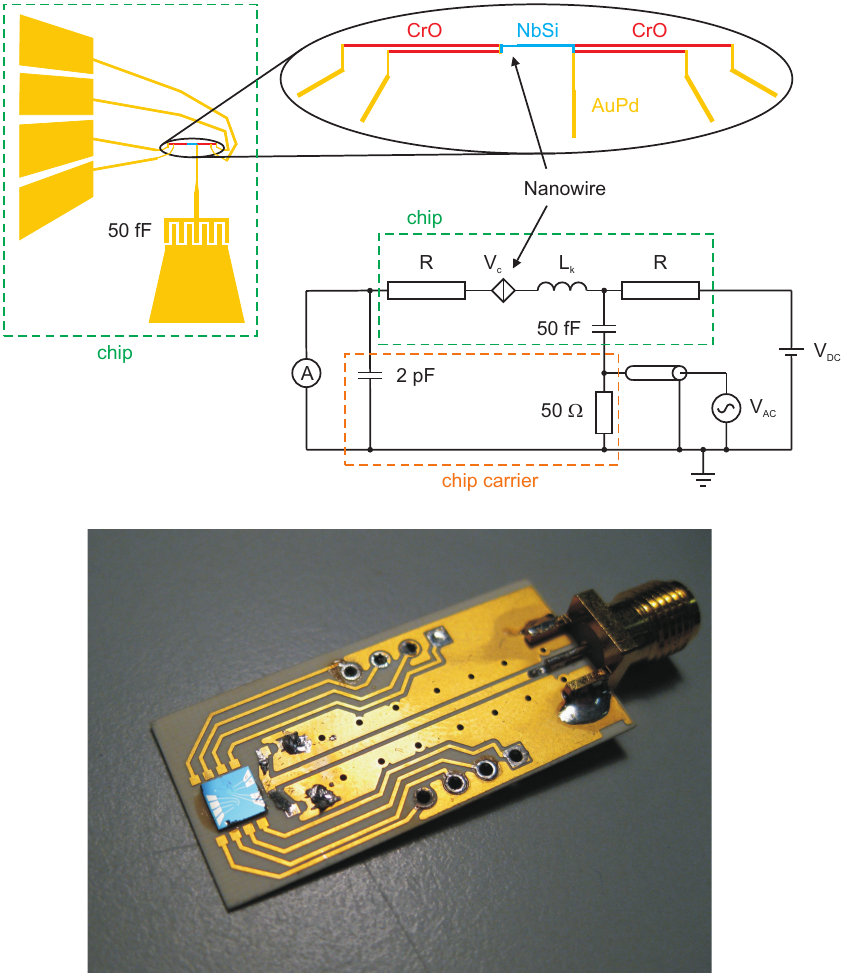}
\caption{Chip layout (not to scale) and circuit schematic.  The
photograph shows the chip mounted on the chip carrier. Note that the
interdigitated capacitor and termination resistors are not present
in the photograph.} \label{fig:QPS-Circuit}
\end{center}
\end{figure}

Figure \ref{fig:QPS-Circuit} shows our superconducting nanowire
sample and its environment. A NbSi nanowire of width 18--19 nm ({\it
cf.} the estimated NbSi coherence length $\approx$20 nm
\cite{Hongisto-2012}), length 5 $\mu$m and thickness 10 nm was
fabricated in series with a wider section of NbSi (width 25 nm,
length 25.2 $\mu$m) which served as an additional kinetic inductance
$L_{\rm k}$ exhibiting negligible QPS rate (since the QPS energy
depends exponentially on the cross-sectional area of the
wire,\cite{Arutyunov-review} a small increase in width results in a
large decrease in QPS energy.)  The nanowire and $L_{\rm k}$ line
were embedded within a high-impedance environment comprising two CrO
resistors each with design dimensions 70 $\mu$m $\times$ 80 nm
(width) $\times$ 10 nm (thickness).  The NbSi and CrO components
were connected via a AuPd micropad interface to ensure good adhesion
and good electrical contact.  AuPd was also used to fabricate the
bond pads and an interdigitated capacitor (design value 50 fF) to
enable ac biasing of the nanowire.

The chip was mounted on a chip carrier containing an SMA launcher
and coplanar waveguide to transmit ac signals to the chip, as well
as bond pads and leads to enable dc biasing.  The waveguide was
terminated with two 100 $\Omega$ resistors in parallel to provide a
termination impedance of 50 $\Omega$.  A 2.2 pF capacitor was
soldered between one of the dc leads and the ground plane to ensure
a low-impedance path for ac signals to return to ground without
passing through the current measurement path.

The NbSi nanowire and the CrO resistors were fabricated by the same
method as described in Ref.~\onlinecite{Hongisto-2012}.  The
substrate was silicon with a 300 nm thermal-oxide layer.  First, the
CrO resistors were fabricated together with AuPd contact wires and
micropads in a single vacuum cycle utilizing shadow evaporation
through a bilayer polymethyl methacrylate (PMMA) and copolymer
stencil mask.  The CrO resistors were evaporated at a low residual
pressure of oxygen ($\sim 10^{-6}$ mbar) followed by a 50 nm layer
of AuPd from an angle for which the narrow stencil openings for the
CrO resistors were overshadowed by the mask.  This avoided the
formation of AuPd shadows parallel to the CrO resistors.  Next, the
CrO resistors and other parts of the circuit were protected by a
PMMA mask while a 10 nm film of amorphous Nb$_x$Si$_{1-x}$ ($x =
0.45$) was co-sputtered on the substrate, making contact only with
the AuPd micropads.  The value of $x$ was defined by first
calibrating the sputter rate for each element at a given sputter
power and periodically confirmed by energy-dispersive x-ray (EDX)
measurement using separately deposited films on Ge substrates.  To
remove organic residue and water before NbSi deposition, the
micropads were cleaned by reactive ion etching with oxygen plasma
and baking overnight in a N$_2$ atmosphere at a temperature of
120$^{\circ}$C. The substrate was rotated during deposition to
ensure homogeneity of the film. After lift-off, the wafer was coated
with inorganic negative-tone hydrogen silsesquioxane (HSQ) resist
(XR-1541, Dow Corning) and patterned with an electron beam to define
the nanowire and $L_{\rm k}$ line. Finally, the sample was etched
with an inductively coupled plasma etching process with SF$_6$ gas.

The sample contained four CrO resistors in total, two of length 70
$\mu$m and two of length 50 $\mu$m (see Fig.\ \ref{fig:QPS-Circuit})
to enable us to measure the resistance of the CrO resistors
independently of the NbSi components.  At room temperature, the
resistance per unit length of the resistors was 9.25
k$\Omega\,\mu$m$^{-1}$ and a single CrO resistor of length 70 $\mu$m
had a resistance of 648 k$\Omega$.  \footnote{More extensive tests
with CrO-resistor-only samples confirmed that the resistances scale
linearly with their dimensions, validating this analysis.}

The resistivity of the NbSi film was measured independently at $T =
3.4$ K (above $T_{\rm c} \approx 1$ K)\footnote{The NbSi film
parameters were measured on stand-alone test wires 20 nm in width.}
and found to be $\rho_{\rm N} =$ 560 $\mu\Omega$cm.
\cite{Hongisto-2012}  Using the design dimensions of the nanowire
and $L_{\rm k}$ line to calculate their normal state resistances we
obtain 165 k$\Omega$ for the nanowire and 565 k$\Omega$ for the
$L_{\rm k}$ line, yielding a total for the NbSi components of 730
k$\Omega$, which agrees well with the measured value of 705
k$\Omega$.  We can estimate the kinetic inductance $L_{\rm K}$ of
the NbSi components at $T = 0$ using the formula\cite{Mattis-1958}
\begin{equation}
L_{\rm K}(0) = \frac{\hbar R_{\rm N}}{\pi\Delta(0)} = 0.18\frac{\hbar R_{\rm N}}{k_{\rm B}T_{\rm c}}.
\end{equation}
This yields $L_{\rm K} \approx 1000$ nH.

\section{dc Measurements}
We mounted the circuit shown in Fig.\ \ref{fig:QPS-Circuit} in a
dilution refrigerator and measured the $IV$ characteristic under dc
voltage bias. The dc lines were passed through feedthrough
$\pi$-filters before entering the cryostat and were subsequently
filtered at 4 K and at base temperature.  The filters at 4 K had an
attenuation of $\sim 90$ dB up to 2 GHz, dropping to $\sim 30$ dB
above 3 GHz.  At base temperature, the copper powder filters had an
attenuation of 30 dB at 40 MHz, rising to $> 90$ dB above 100 MHz.
The chip carrier was mounted within a copper enclosure on the tail
of the dilution refrigerator which was bolted to the mixing chamber.
dc voltage bias was provided by a source meter on the 20 V range,
which had a noise floor of 1 $\mu$V\,Hz$^{-1/2}$ and a spike of
amplitude 55 $\mu$V\,Hz$^{-1/2}$ at 50 Hz.  We used a $\times$ 10000
divider to further reduce the noise amplitude.  dc current was
measured using a low-noise current pre-amplifier with a gain of
$10^8$ V/A, followed by a multimeter.

\begin{figure}[!ht]
\begin{center}
\includegraphics[width=8.6cm]{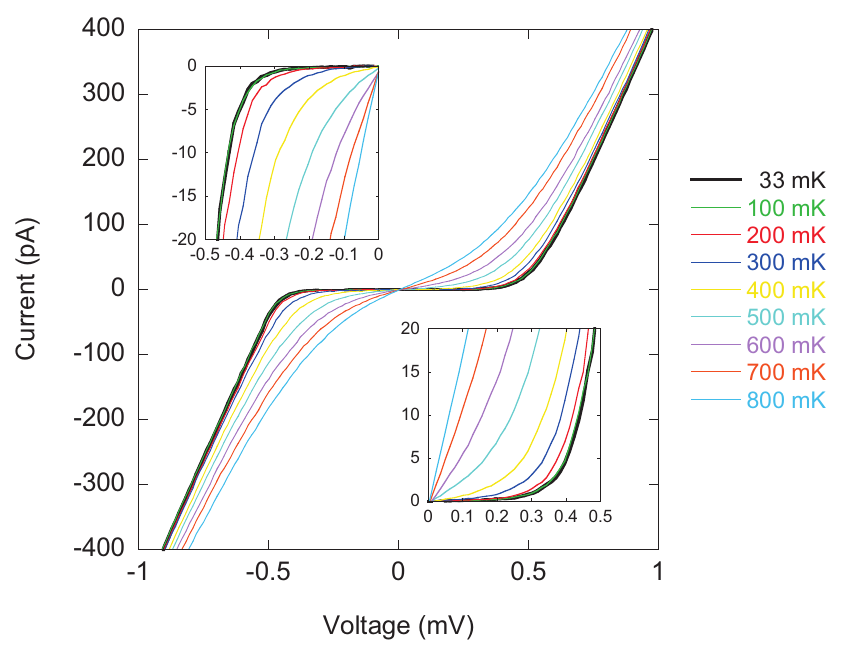}
\caption{$IV$ characteristic of the NbSi nanowire embedded in its
high-impedance environment.  The thick dark line was measured at 33
mK.  The coloured lines show measurements made at higher
temperatures.  The insets provide close-up views of the data.}
\label{fig:IV-vs-T}
\end{center}
\end{figure}

The measured $IV$ characteristic is shown in Fig.\
\ref{fig:IV-vs-T}.  At the base temperature\footnote{The temperature
was recorded by a RuO$_2$ thermometer located at the mixing chamber
of the dilution refrigerator.}, current blockade can be clearly
seen. Controlling the power supplied to the still heater, the
temperature dependence was measured. No change was observed between
33 mK (base temperature) and 100 mK. The increased rounding observed
at the critical voltage is characteristic\cite{Ambegaokar-1969} of
the effect of the Johnson noise generated in the CrO
resistors.\footnote{We neglect Johnson noise from the NbSi nanowire
itself and the $L_{\rm k}$ line, as these are in the superconducting
state.}

\begin{figure}[!ht]
\begin{center}
\includegraphics[width=8.6cm]{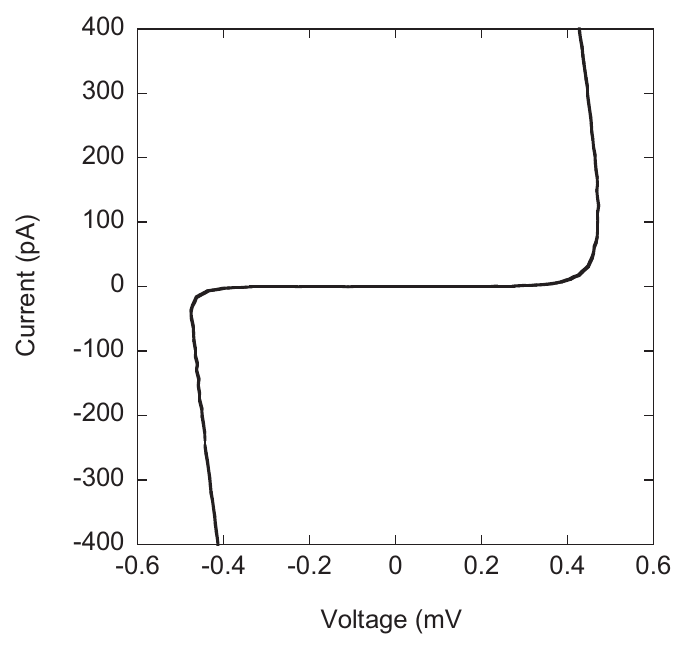}
\caption{$IV$ characteristic at 33 mK with the voltage drop $IR$
across the CrO resistors subtracted.  backbending can be seen in the
transition to the resistive branch, as expected in the presence of
coherent single Cooper-pair tunneling (dual to the coherent QPS
effect\cite{Mooij-2006,Kerman-2012}).} \label{fig:back-bending}
\end{center}
\end{figure}

At first glance, the $IV$ characteristic appears to exhibit an
excess voltage, i.e.~the resistive branch, when extrapolated to zero
voltage bias, intercepts the voltage axis at a non-zero value, which
could be interpreted as being indicative of conventional Coulomb
blockade due to single-electron tunneling.\cite{Fulton-1987}
However, as we show in Sec. \ref{sec:DCmodel}, there is good
qualitative agreement with numerical simulations of a QPS circuit,
suggesting that, at higher bias voltage, the current would approach
$V/R$ as an asymptote. We have made measurements at higher voltages
on other, nominally identical, samples which support this view and
measurements to higher bias on somewhat similar samples in
Ref.~\cite{Hongisto-2012} show no excess voltage. Further, Fig.\
\ref{fig:back-bending} shows the inferred voltage drop across the
superconducting nanowire (by subtracting the voltage drop $IR$
across the CrO series resistors from the measured voltage). There is
backbending in the transition to the resistive state, as expected
for coherent single Cooper-pair tunneling.\cite{Hongisto-2012,
Likharev-1985}

\section{ac Measurements with a continuous-wave signal}
In order to seek dual Shapiro steps as an unambiguous demonstration
of the potential of the nanowire to exhibit QPS, we applied an
additional continuous-wave ac voltage signal while sweeping the dc
voltage bias.  We applied ac signals at various frequencies and here
present representative measurements at two frequencies, 100 MHz and
1.3 GHz. The ac signal was supplied by a high-power swept-signal
generator, and a total of 55 dB of attenuation was included between
the signal generator and the sample.\footnote{Several attenuators
were placed at intervals along the semi-rigid coaxial microwave line
in the dilution fridge: 3 dB at 4 K, 10 dB at the 1 K pot, 10 dB
between the 1 K pot and the mixing chamber and 1 dB at the mixing
chamber. An additional 11 dB of attenuation was present in the line
itself and imperfections in the connections (measured by a vector
network analyzer).  Outside the dilution fridge, an additional 20 dB
attenuator was used to reduce the amplitude of the signal leaving
the source, which had a range of -20 dBm to 0 dBm.}  The ac voltage
amplitude dropped across the 50 $\Omega$ termination resistance on
the chip carrier and, therefore, varied from 40 to 400 $\mu$V rms as
the total ac power varied from $-75$ to $-55$ dBm.

\begin{figure}[!ht]
\begin{center}
\includegraphics[width=8.6cm]{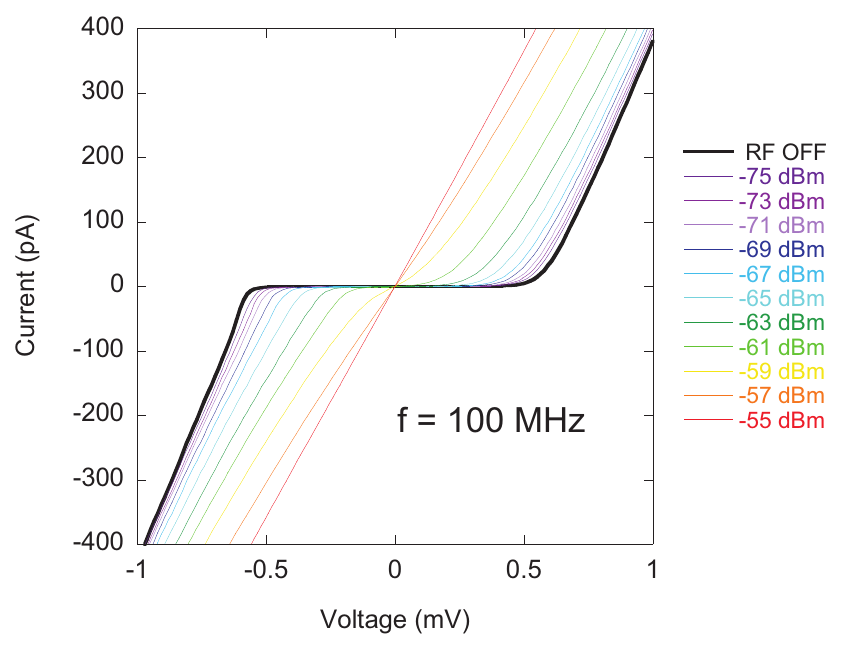}
\caption{$IV$ characteristic under a continuous-wave ac bias of 100
MHz, measured at 33 mK.  The thick dark line is the $IV$ without the
ac bias.  The coloured lines show the gradual change that occurs as
the amplitude of the ac signal is increased in steps of 2 dB from
$-75$ dBm (violet) to $-55$ dBm (red).} \label{fig:IV-vs-CWPower}
\end{center}
\end{figure}

The measured $IV$ characteristic at $f = 100$ MHz is shown in Fig.\
\ref{fig:IV-vs-CWPower}.  As the amplitude of the ac signal is
increased, the blockade voltage becomes less pronounced and more
rounded until at $-55$ dBm the curve is completely ohmic.  No dual
Shapiro steps were observed (expected at multiples of 32 pA at 100
MHz). The increased rounding of the curve with increasing ac
amplitude is similar to the rounding caused by higher temperature
(cf. Fig. \ref{fig:IV-vs-T}) but, at higher bias, the current for a
given bias voltage is much larger than expected due to heating;
these features are similar to features expected in the presence of
QPS, as we show in Sec. \ref{sec:ACModel}.

\begin{figure}[!ht]
\begin{center}
\includegraphics[width=8.6cm]{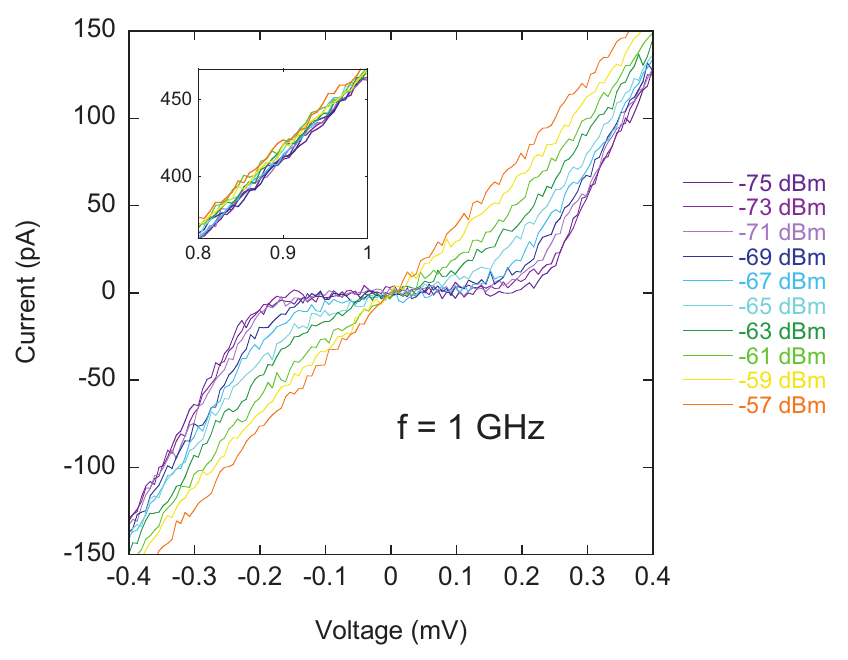}
\caption{$IV$ characteristic under a continuous-wave ac bias of 1.3
GHz, measured at 33 mK for increasing ac amplitudes.  The inset
shows the region in which a dual Shapiro step would be expected (420
pA).  The increased noise present in these data, compared to that
obtained at 100 MHz, is due to the use of a different current
preamplifier and a lower divider of $\times$ 10 on the output of the
voltage source.} \label{fig:IV-vs-CWPower_1-3GHz}
\end{center}
\end{figure}

In our case of large damping, at higher frequencies one expects dual
Shapiro steps to have a larger maximum width $V_n\approx
2(\omega/\omega_{\rm c})V_{\rm c}=(4/2\pi)e\omega{}R$, where
$\omega_{\rm c} = 2\pi V_{\rm c}/2eR$, \cite{Likharev} and,
therefore, to be more robust against the effects of heating.
However, when we applied an ac signal at a higher frequency of 1.3
GHz, we once again observed no dual Shapiro steps (these would be
expected at multiples of 420 pA at 1.3 GHz). Figure
\ref{fig:IV-vs-CWPower_1-3GHz} shows the measured IV characteristic
at 1.3 GHz, applied to a different sample with the same nominal
design. Increased noise present in these data (compared to that
obtained at 100 MHz) is due to the use of a different current
preamplifier and a lower divider of $\times$ 10 on the output of the
voltage source. The transition to the resistive state once again
becomes more rounded at higher amplitudes, consistent with
significant heating.

The observed blockade voltage in this sample was lower (around 250
$\mu$V), although the nanowire dimensions were nominally virtually
identical. This critical-voltage $V_{\rm c}$ variation between
samples is likely to be associated with small inhomogeneities in the
cross section of the wires which result in areas of locally
increased resistance where the QPS rate is enhanced so the value of
$V_{\rm c}$ depends on the number, strength and relative phases of
these spots.\cite{Hongisto-2012,Vanevic-2012}  We note that it was
argued in Ref.~\onlinecite{Hongisto-2012} that the behavior of a
sufficiently short wire containing several weak segments at
different positions is similar to that of a single QPS junction
having equivalent energy $E_{\rm S}$ equal to a (vector) sum of
individual QPS energies. This sum is maximised when the electric
polarization of sections of nanowire between the weak segments is
uniform (or zero), corresponding to equal phases for the individual
QPS segments.

\section{Numerical model of the $IV$ characteristic of a QPSJ}
In order to better understand the effect of an elevated electron
temperature on the $IV$ characteristic of a QPSJ, we have
constructed a numerical model for the circuit dynamics which
includes a temperature-dependent Johnson noise term and in this
section we present simulations of the system at a number of
different temperatures.

\begin{figure}[!ht]
\begin{center}
\includegraphics[width=8.6cm]{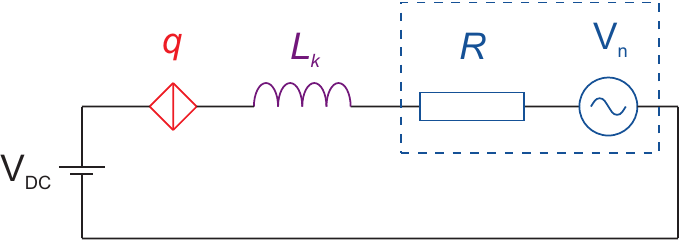}
\caption{Equivalent circuit of a quantum phase-slip junction.  The
resistance $R$ includes both CrO resistors.  Johnson noise is
included by modelling the resistance as an ideal voltage source with
output resistance $R$ and a random, Gaussian signal as a function of
time.} \label{fig:QPSJ}
\end{center}
\end{figure}

A QPSJ consisting of a superconducting nanowire in series with an inductance and a resistance (see Fig.\ \ref{fig:QPSJ}) is the exact dual of a Josephson junction shunted by a capacitor and a resistor (RCSJ).\cite{Mooij-2006}  Therefore, a QPSJ can be modelled using an equation of motion exactly dual to that of the RCSJ model:
\begin{equation}
V(t) = V_{\rm c}\sin\left(\frac{2\pi q}{2e}\right) + L_{\rm K}\frac{\dif ^2 q}{\dif t^2} + R\frac{\dif q}{\dif t} + V_{\rm n},
\label{eq:V(t)}
\end{equation}
where $V(t)$ is the applied voltage bias (dc or a combination of dc
and ac), $q$ is the total charge to pass through the nanowire since
time $t=0$, $e$ is the charge on an electron, $V_{\rm c}$ is the
critical voltage of the nanowire, $L_{\rm K}$ is the combined
kinetic inductance of the nanowire and the wider $L_{\rm k}$ line,
$R$ is the combined resistance of the CrO resistors, and the
Johnson-noise voltage generated by $R$ is given by
\begin{equation}
V_{\rm n,rms} = \sqrt{4k_{\rm B}RTB},
\end{equation}
where $k_{\rm B}$ is Boltzmann's constant, $T$ is the electron temperature and $B$ is the bandwidth of the circuit.

We used a fourth-order Runge-Kutta technique, with a
Gaussian-distributed random-number generator for $V_n(t)$, to solve
this equation for a series of dc voltage bias points $V_{\rm bias}$,
yielding oscillatory solutions for $q(t)$ and $I(t)={\rm d}q/{\rm
d}t$. By averaging $I(t)$ over many cycles, we obtained a value for
the dc current corresponding to each dc voltage bias point, and thus
obtained simulations of the $IV$ characteristic.   None of our
simulated $IV$ curves is completely smooth, due to the finite number
of points over which the dc current was averaged.

\subsection{dc simulations}\label{sec:DCmodel}
For a dc voltage bias, the drive term in Eq.\ \ref{eq:V(t)} is
simply $V(t) = V_{\rm bias}$.  In order to understand the effect of
an elevated electron temperature on the dc $IV$ curve, we choose
simulation parameters that give good qualitative agreement with the
data shown in Fig.\ \ref{fig:IV-vs-T}: $V_{\rm c} = 700$ $\mu$V,
$L_{\rm K} = 890$ nH and $R = 1.6$ M$\Omega$.  These are close to
the design values for the measured circuit: $L_{\rm K}$ = 1000 nH
and $R = 1.3$ M$\Omega$.  We find that the shape of the $IV$ curve
is rather insensitive to the value chosen for $L_{\rm K}$, because
the simulation parameters lie in the strongly damped regime with a
$Q$-value of 0.069.\footnote{For a QPSJ, the $Q$-value is given by
$Q = \sqrt{2\pi V_{\rm c}L_{\rm K}/2eR^2}$.}  The bandwidth of the
circuit below $V_{\rm c}$ is limited by the cutoff frequency,
$\omega_{\rm c} = 1/RC_{\rm QPS,0}\approx 8\times 10^9$ s$^{-1}$,
where $C_{\rm QPS,0} = 2e/(2\pi V_{\rm c})$ is the (zero-voltage)
QPS nonlinear capacitance analogous to the (zero-current) Josephson
inductance $L_{\rm J}=\Phi_0/(2\pi{}I_{\rm c})$.\footnote{Since
$Q\ll 1$, $\omega_{\rm c}\ll \omega_{\rm p}$, where the plasma
frequency $\omega_{\rm p} = 1/\sqrt{L_{\rm K}C_{\rm QPS,0}}$.} The
bandwidth varies as a function of voltage bias below $V_{\rm c}$,
decreasing as the bias voltage increases towards $V_{\rm c}$, due to
the increase of the capacitance to $C_{\rm QPS}=C_{\rm
QPS,0}/\cos{(\arcsin{V/V_c})}$. Above $V_{\rm c}$ the bandwidth is
limited by the roll-off frequency of the series resistors, which
become capacitive at high frequencies. For our simulations, for
simplicity, we chose a fixed bandwidth of 18 GHz, similar to the
roll-off frequency of the CrO resistors.\cite{Zorin-2000}

\begin{figure}[!ht]
\begin{center}
\includegraphics[width=8.6cm]{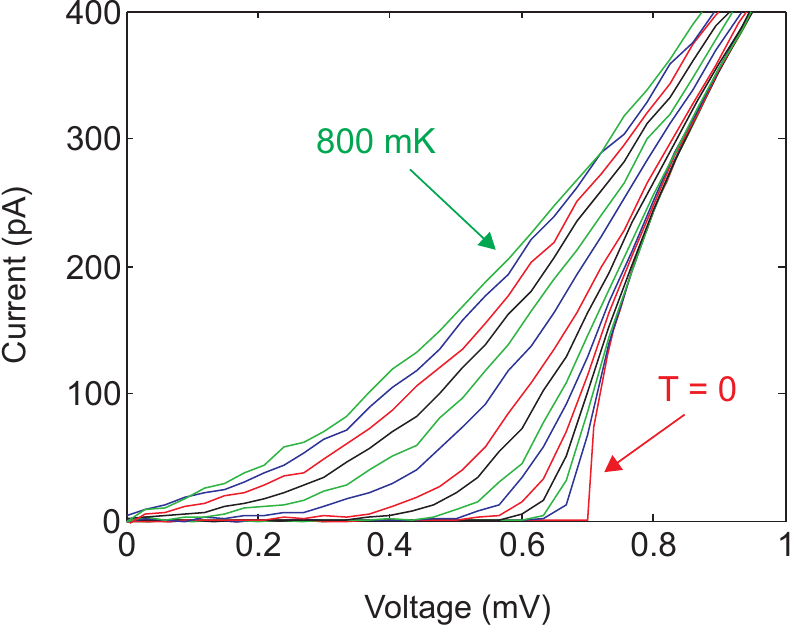}
\caption{Simulated $IV$ characteristic at temperatures 0 K, 10 mK,
20 mK, 33 mK, 50 mK, 75 mK, 100 mK, 150 mK, 200 mK, 300 mK, 400 mK,
500 mK, 600 mK, 700 mK and 800 mK.  The ripples in the curves are
due to the finite number of points over which the dc current was
averaged.} \label{fig:T-Simulations}
\end{center}
\end{figure}

Fig.\ \ref{fig:T-Simulations} shows simulations of the dc $IV$
characteristic at temperatures corresponding to those for which the
curve was measured experimentally.  The simulations are
qualitatively similar to the measured curves shown in Fig.\
\ref{fig:IV-vs-T}, with the best agreement obtained at the highest
temperatures.  There are differences in some details.  For example,
at lower temperatures, the differential conductance in the
simulations first increases with increasing $V$ bias and then above
some voltage $V\lesssim V_c$ decreases as $V$ bias is further
increased, whereas in the experimental data the differential
conductance is increasing as $V$ bias increases from zero up to the
highest bias at which we measured. This difference appears not to be
associated with heating and is likely to relate to details of the
experimental circuit not included in the model for the simulations.
One possible reason for the behaviour could be frequency dependence
of the resistor impedance associated with its parasitic capacitance.

\begin{figure}[!ht]
\begin{center}
\includegraphics[width=8.6cm]{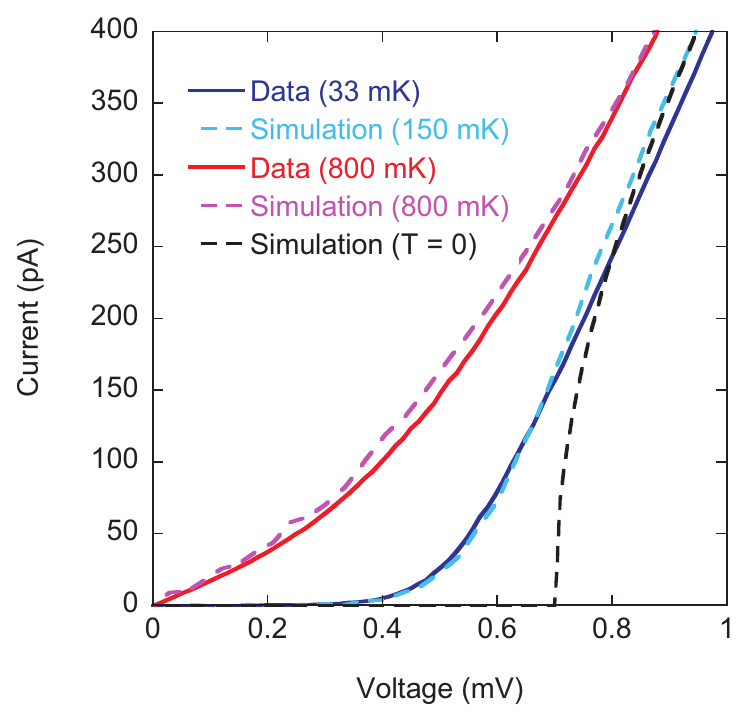}
\caption{Comparison of simulated $IV$ characteristics with data
taken at 33 mK and 800 mK.} \label{fig:T-Simulations+Data}
\end{center}
\end{figure}

Fig.\ \ref{fig:T-Simulations+Data} compares the simulations with the data, showing good qualitative agreement between data and simulation at 800 mK.  The data taken at 33 mK (the base temperature of the dilution refrigerator) agree best with a simulation at 150 mK, suggesting an elevated electron temperature.

\subsection{Continuous-wave ac simulations}\label{sec:ACModel}
\begin{figure}[!ht]
\begin{center}
\includegraphics[width=8.6cm]{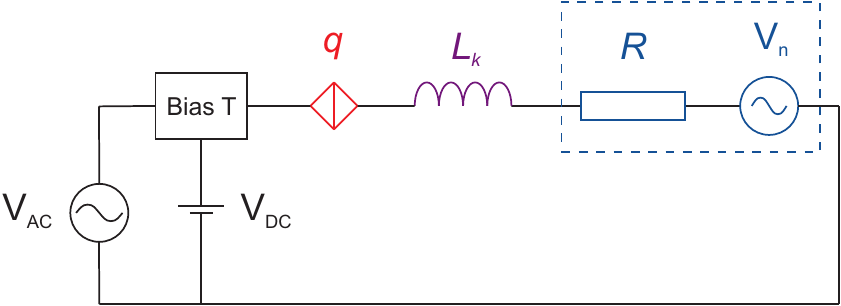}
\caption{Quantum phase-slip junction with combined dc and
continuous-wave ac voltage bias.  The resistance $R$ includes the
resistance of both CrO resistors.} \label{fig:BiasTee}
\end{center}
\end{figure}

To model the behaviour in the presence of an ac bias signal, we
consider a circuit in which the ac and dc bias voltages are combined
via a bias tee before being applied to the QPSJ circuit including
both CrO resistors (Fig.\ \ref{fig:BiasTee}).  The drive term in
Eq.\ \ref{eq:V(t)} then becomes $V(t) = V_{\rm bias} + V_{\rm
ac}\sin\omega t$, where $V_{\rm ac}$ is the amplitude and $\omega =
2\pi f$ is the angular frequency of the ac signal. This circuit is
slightly simpler than the actual circuit shown in Fig.\
\ref{fig:QPS-Circuit}, where the ac currents passing through the two
CrO resistors differ.

\begin{figure}[!ht]
\begin{center}
\includegraphics[width=8.6cm]{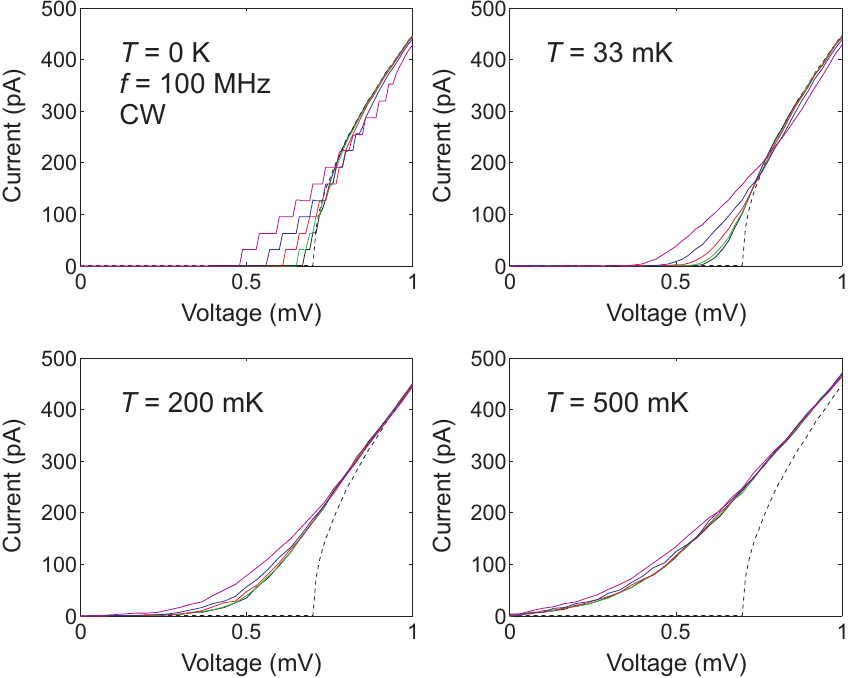}
\caption{Simulated $IV$ characteristics at $T = 0$ K, 33 mK, 200 mK
and 500 mK for a continuous-wave ac signal of frequency 100 MHz and
amplitude 40 $\mu$V (black), 63 $\mu$V (green), 100 $\mu$V (red),
160 $\mu$V (blue) and 250 $\mu$V (purple).  The ripples in the
curves are due to the finite number of points over which the dc
current was averaged.  The $IV$ curve at $T=0$ in the absence of an
ac signal is shown as a black dashed line.}
\label{fig:CW-Simulations-100MHz}
\end{center}
\end{figure}

Figure \ref{fig:CW-Simulations-100MHz} shows the simulation results
for a continuous-wave ac signal of frequency 100 MHz and amplitude
$V_{\rm ac}$ ranging from 40 to 250 $\mu$V.\footnote{The amplitude
values used for the ac simulations do not correspond exactly to
those used for the measurements, as the model assumes a simplified
version of the circuit.}  At $T = 0$, the simulations exhibit dual
Shapiro steps at multiples of 32 pA.  At higher $T$, thermal
fluctuations wash out the steps and, for temperatures of 200 mK and
higher, the features arising in the simulation also arise in a crude
rectification model (not shown) in which the dc current is obtained
from the observed dc $IV$ curve by a simple average over an ac cycle
of the currents corresponding to the instantaneous ac voltages. At
33 mK, some departure from the crude rectification model due to the
nonlinearity in Eq.~\ref{eq:V(t)} is still visible. Comparison with
cw experimental measurements in Fig.\ \ref{fig:IV-vs-CWPower} shows
good qualitative agreement and suggests that the electron
temperature in the lowest-$T$ cw measurements lies below 200 mK.

\begin{figure}[!ht]
\begin{center}
\includegraphics[width=8.6cm]{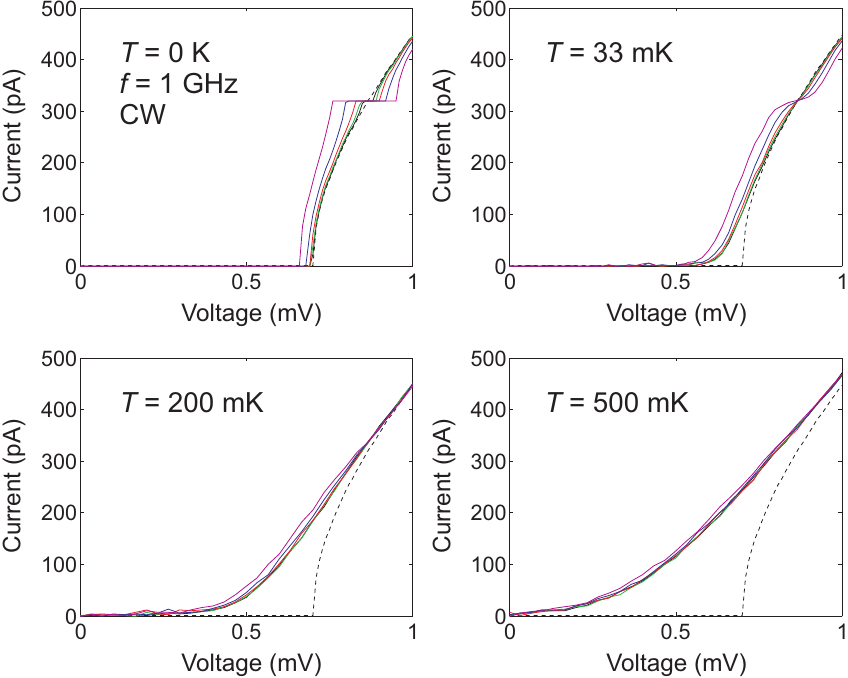}
\caption{Simulated $IV$ characteristics at $T = 0$ K, 33 mK, 200 mK
and 500 mK for a continuous-wave ac signal of frequency 1 GHz and
amplitude 40 $\mu$V (black), 63 $\mu$V (green), 100 $\mu$V (red),
160 $\mu$V (blue) and 250 $\mu$V (purple), applied in series with
the dc voltage bias.  The $IV$ curve at $T=0$ in the absence of an
ac signal is shown as a black dashed line.}
\label{fig:CW-Simulations-1GHz}
\end{center}
\end{figure}

In Fig.\ \ref{fig:CW-Simulations-1GHz}, we show simulations
performed at a higher frequency of 1 GHz, keeping the other
simulation parameters the same as those used at 100 MHz.  At $T =
0$, a single dual Shapiro step appears at a current of 320 pA and
becomes broader as $V_{\rm ac}$ increases.  At $T = 33$ mK, the step
can still be observed, although it is somewhat rounded by thermal
fluctuations. At $T = 200$ mK and higher, the step is washed out.

\subsection{Pulsed simulations}
In the last section we showed that increases in the frequency and
amplitude of the ac signal can be made in order to increase the
width of steps and that this increases the likelihood of observing
steps at higher temperatures. In this section, we show that the step
width can be further increased by applying a pulsed signal, rather
than a continuous-wave signal\cite{Maibaum-2011, Monaco-1990,
Maggi-1996} and once again consider the circuit shown in Fig.\
\ref{fig:BiasTee}. The drive term $V(t)$ now consists of a series of
pulses of amplitude $V_{\rm ac}$, width $w_{\rm p}$ and repetition
rate $\omega/2\pi$, added to the dc voltage bias $V_{\rm bias}$.

\begin{figure}[!ht]
\begin{center}
\includegraphics[width=9cm]{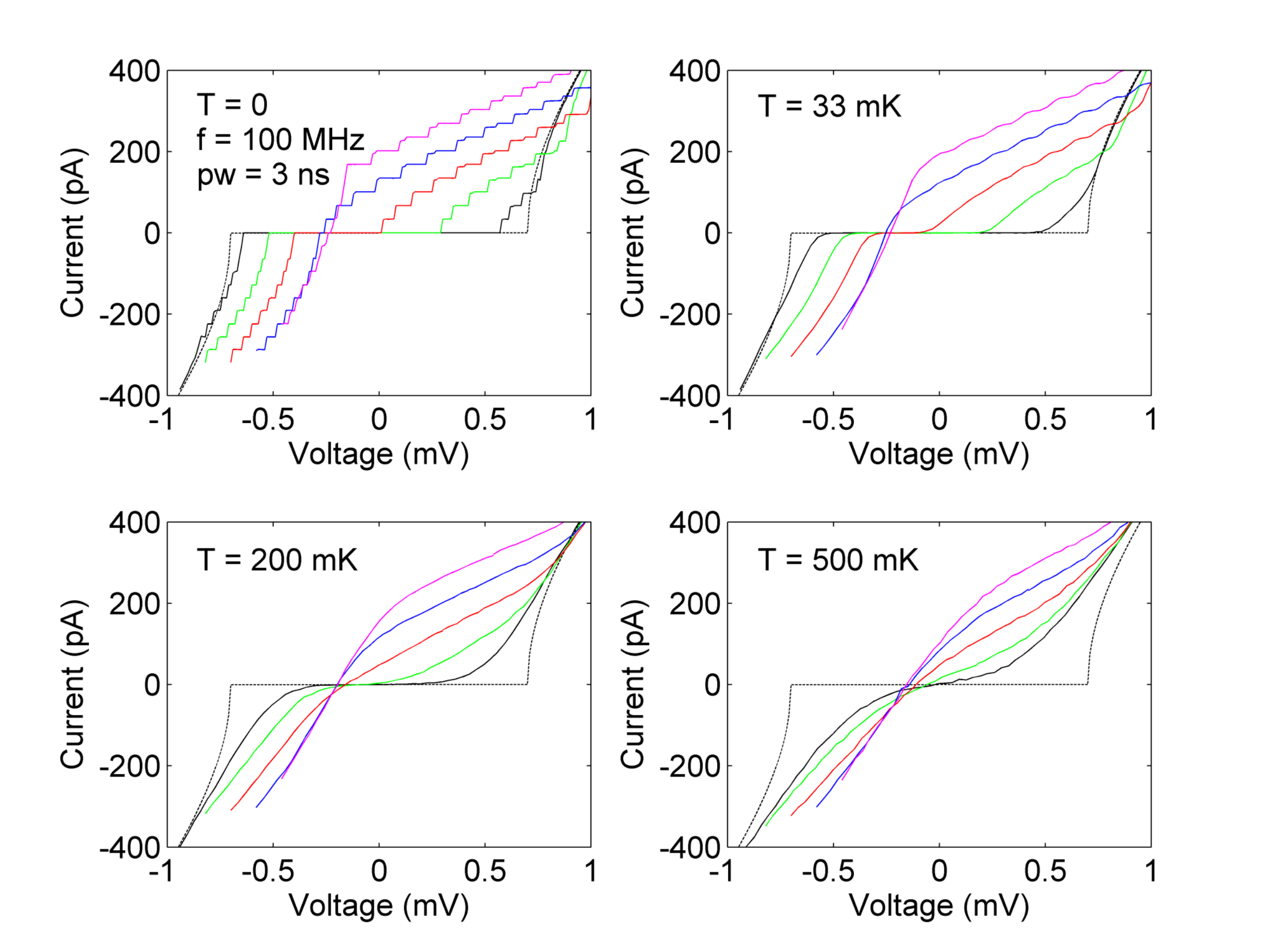}
\caption{Simulated $IV$ characteristics at $T = 0$ K, 33 mK, 200 mK
and 500 mK for a pulsed ac signal with pulse width 3 ns, pulse
repetition rate 100 MHz and amplitude 200 $\mu$V (black), 600 $\mu$V
(green), 1 mV (red), 1.4 mV (blue) and 1.8 mV (purple), applied in
series with the dc voltage bias.  The $IV$ curve at $T=0$ in the
absence of a pulsed signal is shown as a black dashed line.}
\label{fig:Pulsed-Simulations-100MHz}
\end{center}
\end{figure}

Simulations at a pulse repetition rate $f$ of 100 MHz and pulse
width 3 ns are shown in Fig.\ \ref{fig:Pulsed-Simulations-100MHz}
for a variety of pulse amplitudes, keeping the other simulation
parameters the same as for the continuous-wave simulations.  The
first-order dual Shapiro step is expected at $2ef\approx 32$ pA. At
$T = 0$, there are many small dual Shapiro steps, including
noninteger steps due to higher-order processes. At 33 mK, the
integer steps are still visible at the highest pulse amplitudes,
although somewhat rounded. At 200 mK and above, the steps are washed
out.  Compared to the simulations at 100 MHz using a continuous-wave
signal, the steps are observable at a higher temperature.

\begin{figure}[!ht]
\begin{center}
\includegraphics[width=9cm]{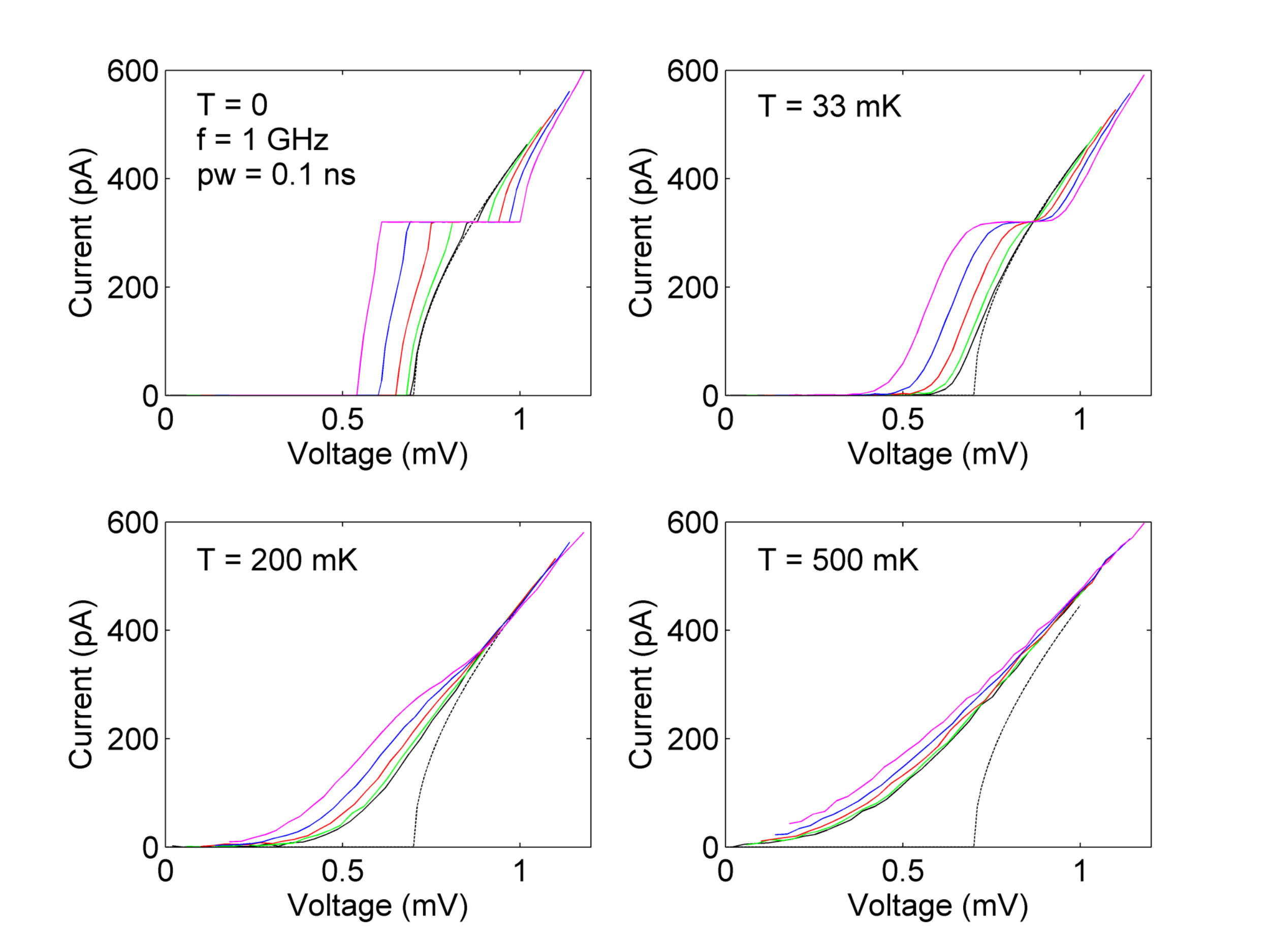}
\caption{Simulated $IV$ characteristics at $T = 0$ K, 33 mK, 200 mK
and 500 mK for a pulsed ac signal with pulse width 0.1 ns, pulse
repetition rate 1 GHz and amplitude 200 $\mu$V (black), 600 $\mu$V
(green), 1 mV (red), 1.4 mV (blue) and 1.8 mV (purple), applied in
series with the dc voltage bias. The $IV$ curve at $T=0$ in the
absence of a pulsed signal is shown as a black dashed line.}
\label{fig:Pulsed-Simulations-1GHz}
\end{center}
\end{figure}

Simulations at a higher pulse repetition rate of 1 GHz and a smaller pulse width of 0.1 ns are shown in Fig.\ \ref{fig:Pulsed-Simulations-1GHz}.  These show a broad dual Shapiro step at $T=0$, the width of which increases with increasing pulse amplitude.  At 33 mK, the step is still visible and exhibits a significant plateau region at the highest pulse amplitudes, despite the rounding due to thermal fluctuations.  At 200 mK, the step is still just visible at the highest pulse amplitudes, although it is washed out at 500 mK.

Therefore, by using a pulsed signal, the step can be made much
broader than for a continuous-wave signal of comparable amplitude.
At 33 mK (the base temperature of our dilution refrigerator), the
step exhibits a plateau of suitable breadth and at a suitable
current ($>$ 100 pA) for a prototype quantum current standard.

\section{Measurements with a pulsed signal}
Since the numerical modeling predicts broader steps for a pulsed
drive than for a continuous-wave ac drive, we carried out
measurements of the $IV$ curve while applying a pulsed signal in
addition to the dc bias voltage.

\begin{figure}[!ht]
\begin{center}
\includegraphics[width=8.6cm]{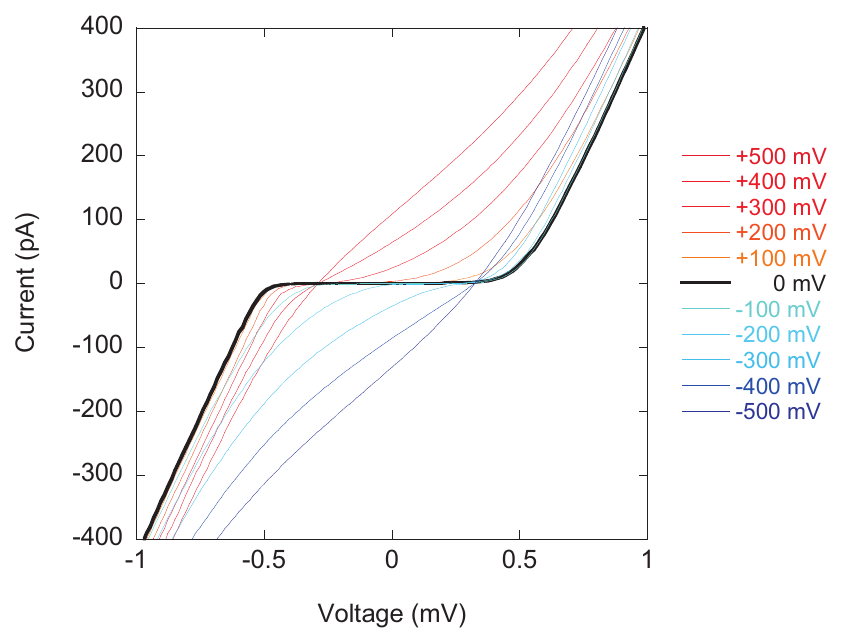}
\caption{Measured $IV$ characteristic using a pulsed drive with a
pulse repetition rate of 100 MHz and pulse width 3 ns, measured at
33 mK.  The thick dark line is the $IV$ curve at a pulse amplitude
of 0 V.  The coloured lines are for pulse amplitudes increasing in
steps of 100 mV in the negative (blue) and positive (red)
directions.  The curve at zero amplitude is not identical to the
$IV$ characteristic when the pulse generator was switched off,
suggesting that some power was being emitted even when the generator
was set to zero amplitude.} \label{fig:IV-vs-PulsedPower}
\end{center}
\end{figure}

Using the same sample as for the continuous-wave ac measurements at
100 MHz, we applied a pulsed signal with repetition rate 100 MHz and
pulse width 3 ns (respectively, the upper frequency limit and lower
pulse-width limit possible with our pulse generator).  Figure
\ref{fig:IV-vs-PulsedPower} shows the measured $IV$ characteristic.
The pulse amplitude at the source ranged from $-500$ mV to $+500$
mV, corresponding to a range of $-2$ to $+2$ mV across the 50
$\Omega$ termination resistance on the chip carrier.

The $IV$ characteristic is not symmetric about the origin, due to
the asymmetric nature of the pulsed signal.  No dual Shapiro steps
are visible.
 Indeed we experimentally probed the $IV$ characteristic over a range of
pulse repetition rates from 10 to 100 MHz and a range of pulse
widths from 3 to 10 ns, but did not observe dual Shapiro steps with
any combination of parameters.

The pulsed experimental data (taken at the base temperature of 33
mK) shows qualitative similarities to the pulsed simulations at 100
MHz at 200 and 500 mK (see Fig.\
\ref{fig:Pulsed-Simulations-100MHz}) which supports the view that
the electron temperature was elevated above the base temperature of
the dilution refrigerator. In the following section we present a
more quantitative evaluation of the influence of heating on the
circuit.

\section{Thermal analysis} \label{sec:thermal}
Comparison of the measurements and simulations presented in the
previous sections point to heating in the circuit washing out any
dual Shapiro steps. Therefore, in this section, we present a simple
lumped-element model of the thermal properties of the QPSJ circuit.
We use this to evaluate the thermal time constants of the system and
to calculate the equilibrium electron temperature in the CrO
resistors. Evaluation of the electron temperature allows us to
consider modifications that can be made to the dimensions and
materials of the resistive elements to reduce the effect of heating.
Evaluation of the time constants allows us to consider the benefit
of pulsed measurements in mitigating heating.

\begin{figure}[!ht]
\begin{center}
\includegraphics{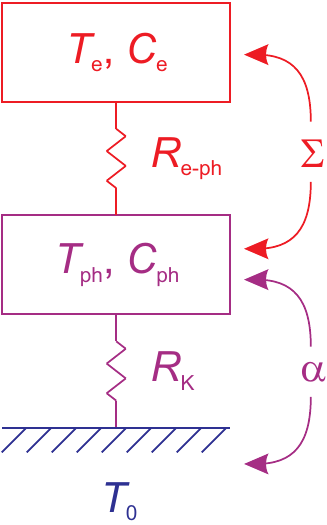}
\caption{Lumped-element model for heat flow from a resistor to the
SiO$_2$ substrate.} \label{fig:Thermal-diagram}
\end{center}
\end{figure}

Our lumped-element model for heat flow in a resistor is shown in
Fig.\ \ref{fig:Thermal-diagram}.  As current is passed through the
resistor, power $P = I^2R$ is dissipated, causing the electron
temperature $T_{\rm e}$ to rise via Joule heating.  Heat flows from
the electron sea to the lattice via the thermal resistance $R_{\rm
e-ph}$ and then from the lattice to the substrate via the Kapitza
thermal boundary resistance $R_{\rm Kapitza}$.  Other relevant
material parameters for the model are $C_{\rm e}$ and $C_{\rm ph}$,
the electronic and phonon specific heat capacities, respectively,
$\Sigma$ is the electron-phonon coupling constant and $\alpha$ is
the thermal coupling constant across the boundary between the
resistor and the substrate.  In what follows, we denote the lattice
temperature $T_{\rm ph}$ and the temperature of the SiO$_2$
substrate, which acts as a cold reservoir, $T_0$.

\subsection{Thermal time constants}
We first consider the response of the system to a change in current.
The electron and lattice temperatures change with time constants
$\tau_1 = R_{\rm e-ph}C_{\rm e}(N_{\rm e}/N_{\rm A})$ and $\tau_2 =
R_{\rm K}C_{\rm ph}(N_{\rm c}/N_{\rm A})$ respectively, where
$N_{\rm A}$ is Avogadro's constant, $N_{\rm e}$ is the total number
of electrons in the resistor and $N_{\rm c}$ is the number of
primitive cells in the lattice.  In the free-electron model, $C_{\rm
e}$ is given by $C_{\rm e} = \gamma T_{\rm e}$, where
\begin{equation}
\gamma = \frac{\pi^2N_{\rm A}k_{\rm B}}{2T_{\rm F}},
\end{equation}
and $T_{\rm F}$ is the Fermi temperature.\cite{Ashcroft+Mermin}
The electron-phonon thermal resistance is given by $R_{\rm e-ph} = 1/5\Sigma\Omega T_{\rm e}^4$.  Inserting these expressions into the expression for $\tau_1$, we arrive at
\begin{equation}
\tau_1 = \frac{n\gamma}{5N_{\rm A}\Sigma T_{\rm e}^3},
\label{eq:tau1}
\end{equation}
where $n = N_{\rm e}/\Omega$ is the electron density.

\begin{table*}[!ht]
\begin{ruledtabular}
\begin{tabular}{|c||c|c||c|c|}
Article & Material & $\Sigma$ (W\,m$^{-3}$\,K$^{-5}$) & Material & $\alpha$ (W\,m$^{-2}$\,K$^{-4}$) \\
\hline
Roukes et al.\cite{Roukes-1985} & Cu & $2 \times 10^9$ & Cu-on-sapphire & 125 \\
Nahum and Martinis\cite{Nahum-1993} & Cu & $3.7 \times 10^9$ & & \\
Verbrugh et al.\cite{Verbrugh-1995} & Al & $0.2$ to $0.5 \times 10^9$ & Al-on-Si & 100 \\
Covington et al.\cite{Covington-2000} & AuPd & $1.4 \times 10^9$ & & \\
\end{tabular}
\end{ruledtabular}
\caption{Values of the electron-phonon coupling constant $\Sigma$
and coefficient $\alpha$ of Kapitza boundary resistance between
resistor and substrate from the literature.} \label{tab:SigmaAlpha}
\end{table*}

We can make a rough estimate of the values of $\tau_1$ and $\tau_2$ for the resistive components as follows.  For Cr, $\gamma = 2.9$ mJ\,mol$^{-1}$\,K$^{-2}$.\cite{Bendick-1982}  For the transition metals Cu, Cd, Ag and Au, $n$ lies in the range 6 to 10 $\times 10^{28}$ m$^{-3}$.\cite{Kittel}  We therefore estimate $n \approx 5 \times 10^{28}$ m$^{-3}$ for CrO.  We estimate the value of $\Sigma$ for both CrO and NbSi to be $1 \times 10^9$ W\,m$^{-3}$\,K$^{-5}$ from measurements made on similar materials (see Table \ref{tab:SigmaAlpha}).

Inserting these values into Eq.\ \ref{eq:tau1} we obtain $\tau_1
\approx 4.8 \times 10^{-8}$K$^3s/T_{\rm e}^3$ for CrO, which gives,
at $T_{\rm e} = 33$ mK, $\tau_1 \approx 1$ ms; at $T_{\rm e} = 200$
mK, $\tau_1 \approx 6$ $\mu$s; and at $T_{\rm e} = 500$ mK, $\tau_1
\approx 400$ ns.  Since these times are all well in excess of pulse
lengths which it is feasible to apply with a repetition rate 100
MHz--1 GHz, temperature variations on the timescale of the pulsed
signal can be neglected.\footnote{Note also that the time step
between dc voltage bias points in the $IV$ curve is much greater
than the time constant for the temperature rise, so after the dc
voltage bias is changed, the electron temperature reaches thermal
equilibrium before the next measurement is made.}

We now evaluate $\tau_2$.  According to the Debye model, the lattice specific heat capacity is given by
\begin{equation} \label{eq:Cph}
C_{\rm ph} = \frac{12\pi^4}{5}N_{\rm A}k_{\rm B}\left(\frac{T_{\rm ph}}{\theta_{\rm D}}\right)^3,
\end{equation}
where $\theta_{\rm D}$ is the Debye
temperature.\cite{Ashcroft+Mermin}  The Kapitza thermal boundary
resistance is given by $R_{\rm Kapitza} = 1/\alpha AT_{\rm ph}^3$.
Now since $\theta_{\rm D} = {\hbar v_{\rm s}}/{k_{\rm
B}}\left({6\pi^2N_{\rm c}}/{\Omega}\right)^{1/3}$, where $v_{\rm s}$
is the velocity of sound, we can use Eq.~\ref{eq:Cph} to write
\begin{equation}
\tau_2 = \frac{2\pi^2}{5}\frac{k_{\rm B}^4}{(\hbar v_{\rm
s})^3}\frac{d}{\alpha}, \label{eq:tau2}
\end{equation}
where $d$ is the thickness of the resistor.  The thickness of the
CrO film is $d = 10$ nm and we take $v_{\rm s}$ in CrO to be 5940
m\,s$^{-1}$.\cite{Samsonov}  We estimate the value of $\alpha$ for
CrO on SiO$_2$ to be 100 W\,m$^{-2}$\,K$^{-4}$ from measurements
made on similar metals and substrates (see Table
\ref{tab:SigmaAlpha}).  Inserting these values into Eq.\
\ref{eq:tau2} we obtain $\tau_2 \approx 60$ ps.  Since $\tau_2$ is
much shorter than $\tau_1$ for all relevant values of $T_{\rm e}$,
any changes in $T_{\rm ph}$ in response to a change in the current passing through the device are for our purposes effectively instantaneous.

\subsection{Electron temperature}
For state-of-the-art parameters for the nanowires (with $V_{\rm c}$
up to 700 $\mu$V), our measurements and simulations indicate that
the electron temperature strongly influences the shape of the $IV$
curve and determines whether it is possible to observe dual Shapiro
steps. In this section, we calculate $T_{\rm e}$ as a function of
the current $I$ passing through the resistor and use the results to
interpret our measurements and simulations.

In thermal equilibrium, the relationship between $T_{\rm e}$ and the
lattice temperature $T_{\rm ph}$ is given by\cite{Roukes-1985}
\begin{equation}
T_{\rm e}^5 = T_{\rm ph}^5 + \frac{I^2 R}{\Sigma\Omega},
\label{eq:TeTph}
\end{equation}
where $\Omega$ is the volume of the resistor.  Similarly, the relationship between $T_{\rm ph}$ and the substrate temperature $T_{\rm 0}$ is given by\cite{Roukes-1985}
\begin{equation}
T_{\rm ph}^4 = T_{\rm 0}^4 + \frac{I^2 R}{\alpha A},
\end{equation}
where $A$ is the surface area of the boundary between the resistor and the substrate.

\begin{figure}[!ht]
\begin{center}
\includegraphics[width=80mm]{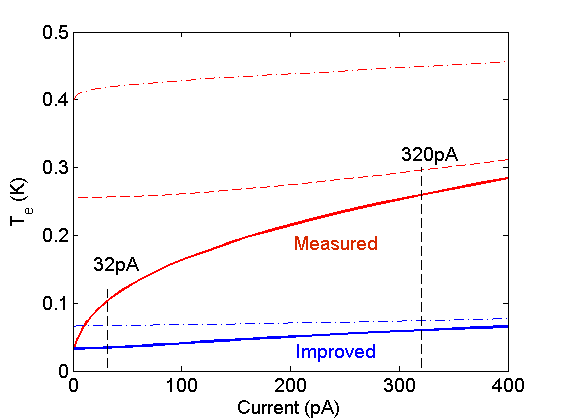}
\caption{Electron temperature $T_{\rm e}$ as a function of current
$I$ for a single CrO resistor  using the dimensions and resistivity
of the measured resistors (solid red line).  For ac bias frequencies
of 100 MHz and 1 GHz, the first dual Shapiro step occurs at 32 pA
and 320 pA, respectively (indicated by vertical black dashed lines).
The dashed red line includes additional heating from a 400 $\mu{}$V
cw signal and the dash-dotted red line includes additional heating
from a superimposed 2 mV pulse signal of duration 3 ns and
repetition rate 100 MHz. The blue curve illustrates the reduction in
$T_{\rm e}$ that can be achieved by increasing the volume and
resistivity of the CrO resistors (see Table
\ref{tab:improved-parameters} for the improved parameters). The
dash-dotted blue line includes additional heating from a
superimposed 5 mV pulse signal of duration 0.1 ns and repetition
rate 100 MHz.}\label{fig:Te_vs_I}
\end{center}
\end{figure}

Using these two equations, we plot the dependence of $T_{\rm e}$ on
the current $I$ for the CrO resistors used in the measured devices
(solid red line in Fig.\ \ref{fig:Te_vs_I}).  The result is almost
identical over the current range probed if we neglect the
temperature difference between the substrate and the lattice.
As the current increases from zero to 400 pA (the range used for
most of the measurements presented in the previous sections), the
electron temperature in the CrO resistors rises from $T_0$ to $280$
mK.  Therefore, unlike in our simulations, an entire measured $IV$
curve is not described by a single electron temperature. Looking
again at Fig.\ \ref{fig:T-Simulations+Data}, we see that, at around
0.65 mV, the curve measured at 33 mK crosses over the simulation at
$T_{\rm e} = 150$ mK.  This is consistent with the experimental
$T_{\rm e}$ lying below 150 mK for $V<0.65$ mV and above 150 mK for
$V>0.65$ mV. At higher $T_0$, the heating due to the applied current
is much less pronounced: for $T_0=800$ mK, at 400 pA, $T_{\rm
e}=801$ mK; the similarity (see Fig.~\ref{fig:T-Simulations+Data})
between the measured and simulated data at this temperature
therefore should be expected.

The application of time-varying signals superposed on the dc V bias
leads to additional heating.  Taking the additional power
dissipation for a continuous-wave ac signal to be $V_{\rm ac}^2/2R$,
for a -55-dBm signal (the largest shown in Figs.\
\ref{fig:IV-vs-CWPower} and \ref{fig:IV-vs-CWPower_1-3GHz}), $V_{\rm
ac}=400$ $µ$V and 0.12 pW of additional power is dissipated in
addition to the dc power dissipation ($\sim$0.25 pW at 32 pA;
$\sim$0.84 pW at 320 pA). The use of a pulsed rather than continuous
signal is well-established as an approach to reducing heating. We
have shown earlier that the use of pulsed signals is beneficial for
obtaining broad steps, and the potentially reduced level of heating
in comparison to cw signals is an additional benefit. Within this
thermal model (for $T_{\rm ph}=T_0$) and when the temperature
variation of $\tau_1$ may be neglected, the temperature rise is
determined by the average power dissipation during a cycle. Based on
the dc IV characteristics, 400-µV pulses of duration 3 ns generate
additional power dissipation $\approx$1 pW for a 100-MHz repetition
frequency, whereas with our improved parameters (see Sec. VIII) much
larger 5-mV pulses of duration 0.1 ns generate additional power
dissipation $\approx$0.2 pW for a 100-MHz repetition frequency or
$\approx$2 pW for a 1-GHz repetition frequency. The additional
temperature rise expected due to the time-varying signals is also
shown in Fig.\ \ref{fig:Te_vs_I} by a dashed red line (for a 400
$\mu$V cw ac signal) and by a dash-dotted red line (for a 2 mV
pulsed signal).

Figure \ref{fig:Te_vs_I} shows that, at a current of 320 pA,
corresponding to the first Shapiro step for a bias frequency of 1
GHz, the electron temperature in the CrO resistors is $260$ mK, even
with no ac signal applied. Our numerical model suggests that, at
these temperatures, the first dual Shapiro step will be washed out.
If dual Shapiro steps are to be observed, the electron temperature
and therefore the amount of heating need to be reduced.

\section{Towards dual Shapiro steps}
In Section \ref{sec:thermal}, we introduced a thermal model for the
system. Changing the dimensions of the resistive elements and/or
their resistivity affects the level of heating. To clarify the
dependence of the electron temperature on the dimensions and the
resistivity of the resistor, we rewrite Eq.\ \ref{eq:TeTph} in terms
of these quantities, taking $T_{\rm ph} = T_0$, $P = I^2R$, $R =
\rho l/wd$ and $\Omega = lwd$, where $\rho$ is the resistivity of
the CrO film, $d$ is the film thickness, $l$ is the length and $w$
is the width of the resistor,
\begin{equation}
T_{\rm e}^5 = T_{\rm 0}^5 + \frac{I^2 \rho}{\Sigma (wd)^2}.
\label{eq:TeRho}
\end{equation}
The electron temperature for a particular current $I$ is dependent
on the resistivity, the width and thickness of the resistor, but
independent of the length.  Since the overall resistance $R$ cannot
be reduced significantly without affecting the dynamics of the
circuit, the best way to reduce $T_{\rm e}$ is by increasing $w$ and
$d$, while simultaneously also increasing the product $\rho l$ (to
keep $R$ unchanged). Increasing $l$ is preferable to increasing
$\rho$, as $T_{\rm e}$ is independent of $l$.  However, three
considerations place upper bounds on $l$: (i) very long resistors
may be less practical to fabricate, (ii) current leakage to ground
via stray capacitance becomes more significant as resistor length
increases\cite{Maibaum-2011}, and (iii) the resistors must be short
in comparison with the wavelength of electromagnetic radiation at
GHz frequencies. The stray capacitance of the resistor is also
affected by changes to the width and thickness, and this determines
the behaviour of the resistor at high frequencies.

The resistivity of the resistors may be modified by changing the
material or its composition, for example by increasing the oxygen
content of the CrO film (although for larger oxygen contents the
resistors become increasingly nonlinear, which is likely to
complicate the behaviour of circuits in which they are
included.\cite{Krupenin-2001}) In Fig.\ \ref{fig:Te_vs_I}, we plot
as a solid blue line $T_{\rm e}$ as a function of $I$ for an
improved set of properties, detailed in Table
\ref{tab:improved-parameters}. The resistivity and all three
dimensions are increased.  We estimate that these changes will
increase the total capacitance of each resistor by a factor of 60,
and that this will cause a corresponding reduction by a factor of 60
in the roll-off frequency of the impedance.  Since the roll-off
frequency of the measured resistors is estimated to be tens of
GHz,\cite{Zorin-2000} the roll-off frequency of the larger proposed
resistors will still be in the GHz range.  With the improved
parameters, for a dc current of 320 pA, the current corresponding to
the first dual Shapiro step for a bias frequency of 1 GHz,
Eq.~\ref{eq:TeRho} predicts an electron temperature in the CrO
resistors of $61$ mK.

\begin{figure}[!ht]
\begin{center}
\includegraphics[width=9cm]{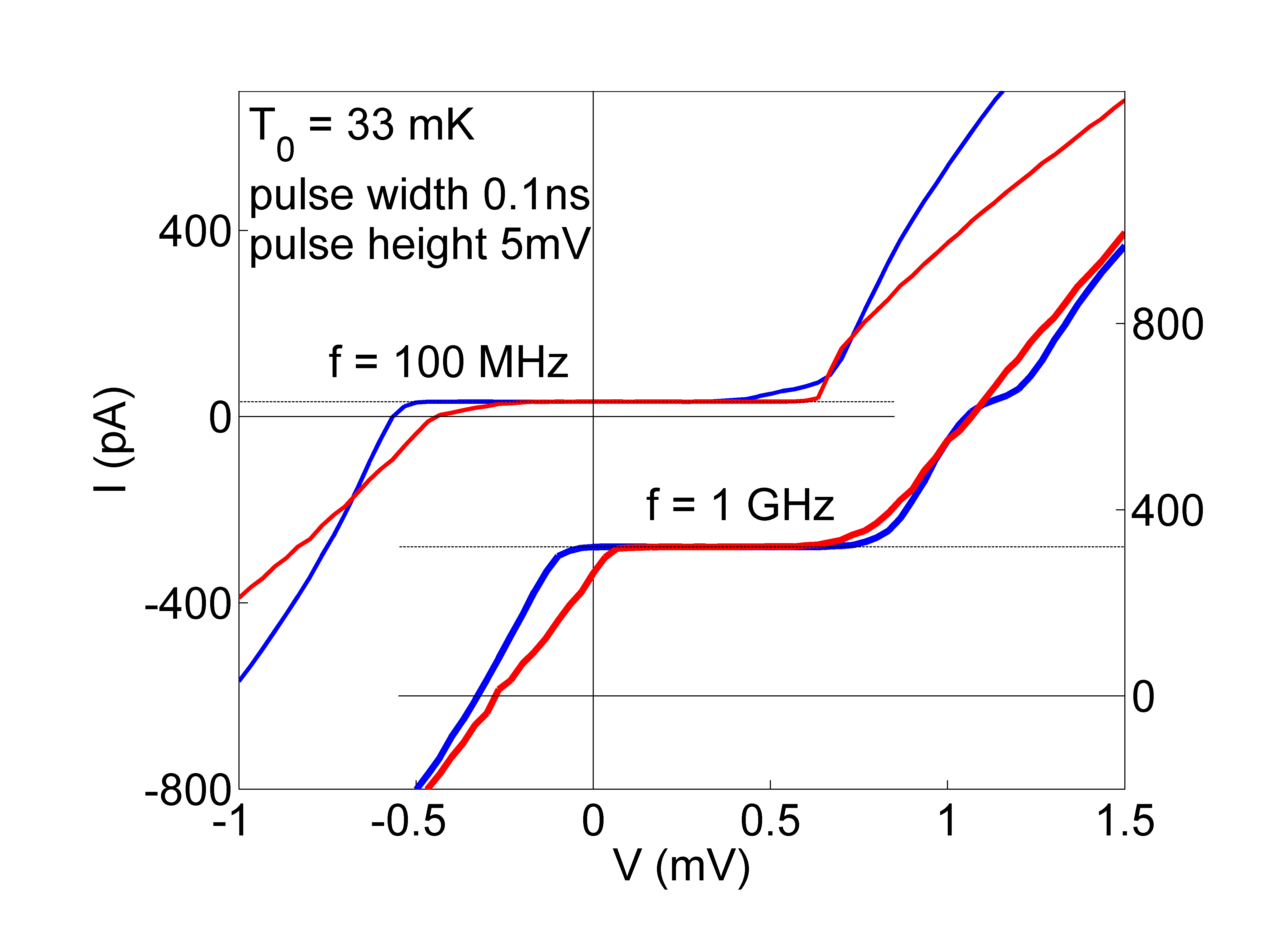}
\caption{Simulated $IV$ curves for a pulsed drive signal of
repetition rate 1 GHz (thick lines) or 100 MHz (thin lines), pulse
width 0.1 ns and pulse amplitude 5 mV for the parameters
corresponding to the measured device (red) and the improved set of
parameters (blue). The simulations assume a substrate temperature
$T_0 = 33$ mK and take into account the dependence of the electron
temperature $T_{\rm e}$ on the current passing through the device.
Green dashed lines show the expected currents at which dual Shapiro
steps are expected. The step is clearly flatter for the improved
parameters.} \label{fig:final-prediction}
\end{center}
\end{figure}

\begin{table}[!ht]
\begin{ruledtabular}
\begin{tabular}{|c||c|c|c|c|}
&$\rho$ (m$\Omega$ cm) & $d$ (nm) & $w$ (nm) & $l$ ($\mu$m) \\
\hline
Measured & 0.74 & 10 & 80 & 70 \\
Improved & 74 & 160 & 2000 & 280 \\
\end{tabular}
\end{ruledtabular}
\caption{Parameters of CrO resistors in the measured device, compared with an improved set designed to reduce heating.}
\label{tab:improved-parameters}
\end{table}
Using our improved parameters for the CrO resistors, we repeated our
simulation of the $IV$ curve for the pulsed drive signal of
repetition rate 1 GHz and pulse width 0.1 ns (thick lines in Fig.\
\ref{fig:final-prediction}), this time including the current
dependence of the electron temperature and iterating to obtain
self-consistent values for the electron temperature and resulting dc
current.

The shorter pulses result in significantly less heating in the CrO
resistors, such that the plateau has sufficient flatness for a
prototype current standard, even with the original resistor
parameters of the measured device.  We found that the amplitude of
the dual Shapiro step generated was maximised by using a higher
amplitude of 5 mV for the pulsed drive signal. This step has the
satisfactory property of crossing the current axis at zero voltage
bias, an ideal property for a current standard.

For a quantum current standard, the current output should exceed 100
pA. At a pulse repetition rate of 1 GHz, the current on the plateau
is 320 pA, fulfilling this requirement. At a slower pulse repetition
rate of 100 MHz with the same pulse width of 0.1 ns, the breadth of
the step is still significant, while the value of the current on the
plateau reduced to 32 pA (thin lines in Fig.\
\ref{fig:final-prediction}.) In this case, an output current
exceeding 100 pA could be achieved with four nanowires in parallel,
all driven by the same pulsed signal.

\section{Conclusion}
We have presented measurements of the voltage-biased $IV$
characteristic of a QPSJ circuit consisting of a NbSi nanowire
embedded in a resistive environment.  The dc $IV$ characteristic
exhibits blockade of the supercurrent and backbending of the
transition to the resistive state, both of which are characteristic
of coherent transfer of single Cooper pairs. We have also reported
the first measurements of the $IV$ curve with an additional ac
voltage bias; we did not find any direct evidence of dual Shapiro
steps either for continuous wave or pulsed signals.

We constructed a numerical model of the $IV$ curve based on QPS,
which included the effect of thermal fluctuations via Johnson noise
in the series resistors.  The simulations were conducted with a
fixed electron temperature but nevertheless provided a good
qualitative fit to the data and enabled us to predict improved
measurement parameters, i.e. a pulsed drive signal with frequency of
order 1 GHz and pulse width of order 0.1 ns, to increase the breadth
of the dual Shapiro steps and hence their robustness to thermal
fluctuations.

In a thermal analysis we considered the time constants of the system
and the variation of the electron temperature with current flow in
the circuit and infer the presence of significant heating in the
measurements. The lack of temperature dependence of the dc $IV$
characteristic for $T<100$ mK, in contrast with the simulation
results in which the dc $IV$ characteristic remains $T$-dependent
below 100 mK, gives further evidence that the electron temperature
of the device was substantially elevated above the substrate
temperature. The result of this heating was sufficient Johnson noise
to wash out any dual Shapiro steps.

Following the thermal analysis, we have predicted an improved set of
parameters for the series resistors to further reduce heating. The
improved parameters would involve resistor composition leading to
highly nonohmic properties and the behaviour of circuits containing
such resistors requires experimental investigation. Our numerical
model predicts that, using these parameters with an increased
amplitude for the pulsed drive signal, an observable Shapiro step
will be observable even in the presence of the heating due to the
drive signal. The step is also predicted to cross the current axis
at zero voltage bias, which would be ideal for a prototype QPS-based
current standard.  In addition we found that, when the repetition
rate of the pulsed drive signal is reduced, the breadth of the step
may be preserved by preserving the short pulse-duration. The lower
repetition rate would have the advantage of reducing
heating\cite{Maibaum-2011} in the series resistors and several
nanowires could be produced in parallel to generate sufficient
current for a prototype current standard.

Our conclusion is that, for any future experiments seeking to detect
dual Shapiro steps, it will be important to make a number of
improvements compared with the experimental set-up which we have
described: (i) increasing the dimensions the series resistors, (ii)
increasing the resistivity of the resistor material used, (iii)
using a pulsed drive signal with a pulse repetition rate of 100 MHz
to 1 GHz, (iv) increasing the pulse amplitude, and (v) reducing the
pulse width to around 0.1 ns.

\begin{acknowledgments}
This research was supported by the UK Department for Business,
Innovation and Skills, the European Metrology Research Programme
(grant no. 217257) and the UK EPSRC.  We also thank Jonathan
Williams and Alexander Tzalenchuk for helpful discussions.
\end{acknowledgments}

\bibliography{QPS}
\end{document}